# Free-space-coupled superconducting nanowire single-photon detectors for infrared optical communications

Francesco Bellei[1], Alyssa P. Cartwright[1], Adam N. McCaughan[1], Andrew E. Dane[1], Faraz Najafi[1], Quinyuan Zhao[1], Karl K. Berggren[1,*]

[1]*Department of Electrical Engineering and Computer Science, Massachusetts Institute of Technology, 77 Massachusetts Avenue, MA, Cambridge,02319, USA*
[*]*berggren@mit.edu*

**Abstract:** This paper describes the construction of a cryostat and an optical system with a free-space coupling efficiency of 56.5% ± 3.4% to a superconducting nanowire single-photon detector (SNSPD) for infrared quantum communication and spectrum analysis. A 1K pot decreases the base temperature to $T = 1.7$ K from the 2.9 K reached by the cold head cooled by a pulse-tube cryocooler. The minimum spot size coupled to the detector chip was $6.6 \pm 0.11$ μm starting from a fiber source at wavelength, $\lambda = 1.55$ μm. We demonstrated efficient photon counting on a detector with an $8 \times 7.3$ μm$^2$ area. We measured a dark count rate of $95 \pm 3.35$ kcps and a system detection efficiency of $1.64\% \pm 0.13\%$. We explain the key steps that are required to further improve the coupling efficiency.

**1. Introduction**

Free-space quantum optical communication in the mid-infrared (mid-IR) [1] is an important technology for applications such as naval operations that cannot rely on optical fibers. The mid-IR range is particularly interesting because of a window of transmission in the atmosphere at wavelength $\lambda = 10$ μm. These communications require high-speed single-photon detectors sensitive to mid-IR radiation. At present, superconducting nanowire single-photon detectors (SNSPDs) [2,3] represent one of the best detectors for this application, due to their single-photon sensitivity, their high speed (few-ns reset time), and their high time resolution (few tens of ps time-jitter). System detection efficiencies

(*SDE*) greater than 67% [4-6] were demonstrated for SNSPDs at near-infrared (near-IR) wavelength ($\lambda = 1.55$ μm). This demonstration was achieved not by increasing the devices detection efficiency but by maximizing the coupling efficiency, i.e. the fraction of photons emitted by the source that are coupled to the SNSPD. In all the cited cases, an optical fiber was aligned to an SNSPD, either passively [4] or actively [5]. We identified three main reasons why we cannot apply the same alignment method to mid-IR optical communications. First, the dimensions of the detector depend on the fiber used. An optical fiber has a fixed core diameter that consequently limits the minimum diameter of the beam emitted by the fiber. To ensure high coupling efficiency, the minimum dimension of the detector has to be larger than the beam diameter; thus, this dimension is also limited. This requirement becomes a non-trivial issue for experiments in the mid-IR, which require optical fibers with a mode-field diameter $MFD > 20$ μm versus the typical active area of an SNSPD which is $\leq 15 \times 15$ μm$^2$. Second, using mid-IR fibers would limit the scalability of the system to multiple channels. Every channel in the system requires an optical fiber; thus, for a large array of detectors a bundle of fibers is required, and integrating this bundle into a cryostat with tight spaces can make the design challenging. Finally, mid-IR optical fibers are more rigid and fragile than near-IR fibers: there is a long-term bend radius of $> 40$ mm for mid-IR fibers vs a long-term bend radius of $> 13$ mm for near-IR fibers. Thus, it would be more difficult to thermalize, cleave, or integrate a mid-IR fiber with components affected by thermal contraction.

An optical system based on free-space optics solves these three problems. In diffraction optics, an optical beam can be focused on a spot whose minimum diameter is smaller than the wavelength of the beam itself. In addition, a free-space optical system can host several channels in a single optical path as long as there is no cross-talk at the receiver. Depending on the spot dimension achieved and the diffraction limit of the system, we can determine how many channels in parallel the system can accommodate.

To date we could not find any demonstration of a cryogenic system with free-space high-coupling efficiency to a single-photon detector in the near- or mid-IR range. Free-space coupling to cryogenic detectors was proposed in the past for astronomical imaging [7] and spectroscopy [8], and for quantum communications [9,10]. Some of these proposals used semiconductor devices. These detectors could be fabricated in arrays with an area four orders of magnitude larger than an SNSPD; however, they were not single-photon sensitive. In addition, the base temperature needed for these devices to operate (~6 K [7], ~50 K [8]), allowed the use of cryostats with a cooling power that is not available below 4.2 K. Other systems used cryogenic detectors to receive quasi-optical millimeter and sub-millimeter radiation [9]. In this case, it is possible to build in-plane antennas that can efficiently focus the radiation onto the detector. Thus, the active area of the receiver is effectively $mm^2$-scale. Verevkin *et al.* [10] used free-space coupling for SNSPDs, but they state that "the working area of our detectors is always smaller and often much smaller than the incident photon beam size". We propose here a cryogenic set-up for superconducting single-photon detectors built to achieve high-efficiency (> 50%) free-space coupling.

We designed and built a vibration-isolating cryostat with free-space optical access able to reach a base temperature $T = 2.9$ K, with an additional stage that can cycle the sample stage to $T = 1.7$ K for 1.5 hours; we measured vibrations amplitudes of $498 \pm 98$ nm at the sample stage. The optical system, composed of two lenses (see Fig. 1(b)), was able to focus light on a detector with a minimum spot waist of $6.6 \pm 0.11$ μm at $\lambda = 1.55$ μm. One of the two lenses was mounted and thermalized inside the cryostat, and it was aligned to the SNSPD chip before the cooldown.

We used an $8 \times 7.3$ μm$^2$ area NbN-on-sapphire SNSPD based on 100-nm-wide nanowires without an optical cavity to calibrate our set-up. We biased the detector at 97% of its switching current (the current at which the detector stops being superconducting and switches to a resistive state). At this set point, we measured a dark count rate of $95 \pm 3.35$ kcps. At the same bias current we measured an *SDE* of $1.64\% \pm 0.13\%$. By characterizing the dimension of the beam at the detector, we estimated a coupling efficiency of $56.5\% \pm 3.4\%$. From the ratio between the two efficiencies, we calculated that the SNSPD's device detection efficiency (*DDE*) was 2.9% at the same bias current.

This document is divided as follows: in Sections 2 and 3, we present the design of the experimental set-up divided into the optical system and the cryogenic system, respectively. In Section 4, we present the results of the mechanical vibrations measurements. In Section 5, we analyze the results obtained from the measurements of *SDE* on large area SNSPDs. Finally, we conclude by discussing the impact of this report on future work.

## 2. Optical system

We designed a three-lens optical system to image the surface of the SNSPD chip and to focus a test beam from a laser source on a detector. A picture and schematic of the system are shown in Fig. 1. The imaging system was designed to achieve a field of view of ~200 µm. Imaging the SNSPD chip allowed us to align the optical beam to the detector and to place Lens 3 at the correct focal distance. Even though the ultimate goal was to use the optical system at mid-IR wavelengths, for this first demonstration we created an optical system able to focus a beam spot of 12.7 µm in diameter at $\lambda = 1.55$ µm.

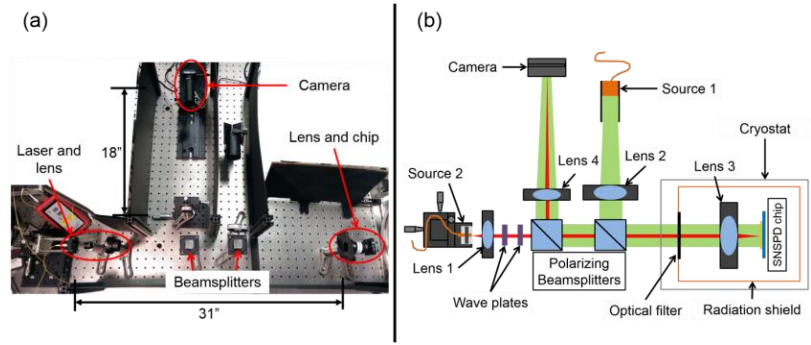

Fig. 1. (a) Picture of the complete optical set-up used for imaging the chip and focusing the light on the SNSPD. The SNSPD chip and Lens 3 were mounted inside the cryostat. The picture was selectively cropped. (b) Schematic of the optical set-up used for chip imaging and beam focusing. The green lines represents the path of the light beam used by the imaging system. The red lines represents the path of the light beam that is focused on the detector. The polarizing beamsplitters are used only for the imaging system. For the detection efficiency

characterization, we use a quarter-wave and a half-wave plate to maximize the transmission through the beamsplitters.

We designed a two-lens telescope to couple > 90% of the light coming from the source onto the active area of the detector; because of the large numerical aperture required to couple $\lambda = 10$ µm light on SNSPD, we had to mount Lens 3 inside the cryostat. The yield in the fabrication process and the constraints in the detector's speed due to the kinetic inductance of the NbN [11], [12], limit our SNSPDs maximum dimensions. An NbN SNSPD with a 15×15 µm$^2$ active area, 80-nm-wide nanowires and 40% fill factor typically showed a reset time of 9 ns, which is as high as we wanted to go for 100 MHz optical communications. We determined from Gaussian optics that in order to have 90% of the source power impinging on an active area of 15×15 µm$^2$, the beam waist had to be no larger than 7.5 µm. As a consequence, we determined that at $\lambda = 10$ µm we needed a numerical aperture of $NA = 0.41$ at Lens 3; thus, for a 25-mm-diameter lens, rather than a larger lens, the maximum acceptable focal length is $f_{Lens3}$ ~28 mm. We chose a 25-mm-diameter lens because increasing the aperture of the optical system not only would have increased the stray light impinging on the detector, but it would have also compromised the cooling ability of the cryostat, due to the increased incoming thermal radiation. In addition, because of the short focal length of Lens 3, we mounted it inside the cryostat (as shown in Fig. 1b). From the commercially available lenses, we selected a C-coated aspheric lens for Lens 3 with $f = 20$ mm.

Starting from our selection for Lens 3 and design wavelength, we picked Lens 1 from commercially available lenses. As we mentioned earlier, for the test described in this document we used a fiber-coupled coherent light source at $\lambda = 1.55$ µm, as a signal. We characterized the beam profile at the output of the optical fiber, and we observed a Gaussian beam with a beam quality $M^2 = 1.35$ and a beam waist of $8.3 \pm 0.05$ µm. Thus, we chose an aspheric lens with $f_{Lens1} = 26$ mm, so that the demagnification of the telescope was 1.3×. In Fig. 2a, we show the profile of the Gaussian beam at the output of the telescope characterized with a beam profiler. We obtained a minimum Gaussian beam waist of 4.82 ± 0.04 µm, which was close to the 4.8 µm waist that we obtained from theoretical calculations [13].

The $f_{Lens4}$ was selected to obtain a field of view of 200 µm. In Fig. 1b, the green lines represent the path followed by the incoherent visible light ($\lambda$ = 635 nm) in the imaging system. The light back-reflected by the SNSPD chip is focused through a telescope formed by lenses 2 and 3 on a CCD camera. The active area of the camera chip is 12.5 × 12.5 mm²; thus, for a field of view of 200 µm we needed a magnification of 25×. Because $f_{Lens3}$ = 20 mm, we used $f_{Lens4}$ = 500 mm. Fig. 2b shows an image acquired with the optical set-up and centered on an SNSPD.

Because of the imaging system, we are able to align the focusing system to the SNSPD in two separate steps [14]. First, Lens 3 was aligned to center the image on the selected SNSPD. In a second step, we aligned the optical source and Lens 1 so that the beam spot was centered on the detector. The use of two separate optical systems allowed us also to verify that Lens 1 was at the focal distance from the SNSPD chip; in particular, we were able to verify that the image was focused on the same plane where the beam from Source 2 was focused.

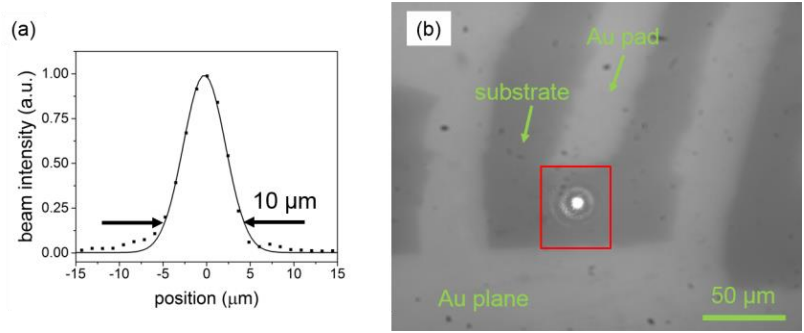

Fig. 2. (a) Beam profile of the spot light focused by the focusing system measured with a beam profiler. The profile is fitted with a Gaussian profile to extrapolate the beam waist. (b) Image of the SNSPD detector on a 200-µm-diameter field of view taken with the optical set-up. The bright dot is the beam from a $\lambda$ = 635 nm laser focused on the chip. The spotlight was moved to place it on the area where an SNSPD was fabricated (inside the red circle).

**3. Cryogenic Set-up**

For single-photon optical communications in the mid-IR, we needed a cryostat able to reach a base temperature $T < 2$ K, even with an optical opening. The two main reasons for needing that temperature are dark counts and material used for the SNSPDs. It has been demonstrated in the past that reducing the base temperature of an SNSPD even by a few tenths of K can significantly change the detector's dark count rate [15]; thus, the temperature should as low as possible. In addition, we wanted to be able to operate SNSPDs based on WSi [16] and not NbN, which are more sensitive to mid-IR photons.

We built a cryostat able to reach a base temperature $T = 1.7$ K at the sample stage; the system was precooled to $T = 2.9$ K with a Cryomech pulse-tube cryocooler (PT415), and then reached base temperature $T = 1.7$ K using a sorption fridge from PhotonSpot Inc. The system was also designed to isolate the mechanical vibrations generated by the pulse-tube from the sample stage (see Section 4).

In Fig. 3a, we show a schematic of the cryostat. The system inside the external chassis can be physically separated into two assemblies. The top assembly is shown in Fig. 3b, and it is mainly responsible for cooling the system: the pulse-tube cryocooler (Cryomech PT415) has a first stage (Stage 1) with a cooling power $P_c = 36$ W at $T = 45$ K, and a second stage (Stage 2) with $P_c = 1.5$ W at $T = 4.2$ K; the cold head of the sorption fridge (PhotonSpot Freeze 4) can reach a temperature $T < 1$ K, for a time that depends on the heat load ($Q_c$) incident on it and on the base temperature reached by the pulse-tube cryocooler. The bottom assembly is shown in Fig. 3c. This assembly hosts the SNSPD chip, and is mounted directly on the optical table; thus, we can perform optical alignment to the detector before cooling down the entire system. The components in Fig. 3b and 3c with the same label have been thermally connected with oxygen-free high thermal conductivity (OFHC) copper. The copper parts are bolted together, with a layer of Apiezon vacuum grease in between to improve the heat conduction. This flexible thermal coupling allowed us to detach the bottom and the top assemblies, and easily access the cooling stages of the system while keeping the optics in the cryostat on the optical table.

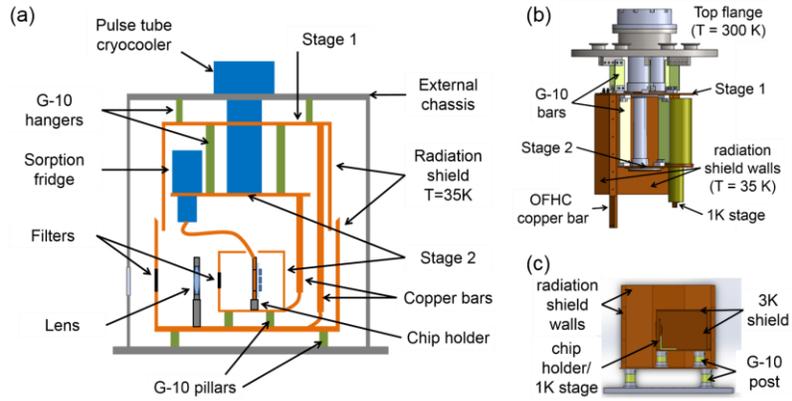

Fig. 3. (a) Schematic of the cryostat. The entire system is also enclosed in a stainless steel chassis. (b) CAD design of the top assembly of the cryostat. (c) CAD design of the bottom assembly of the cryostat.

The materials and the geometry of the system were selected with the goal to minimize $Q_c$ on the pulse-tube cryocooler and on the sorption fridge [17]. Oxygen-free high-conductivity (OFHC) copper is one of the most widely used materials for cryogenic thermal connections because of its unmatched thermal conductivity. G-10 is another widely used material in cryogenics; because of its low thermal conductivity and high rigidity, G-10 is an ideal material for rigid connections between parts at different temperatures.

Table 1 shows the heat load ($Q_c$) budget for the top and bottom assemblies. For the top assembly, as shown in Fig. 3b, we used G-10 bars to hang the stage at $T = 35$ K (Stage 1) from the top flange of the cryostat ($T = 300$ K), and the stage at $T = 3$ K (Stage 2) from Stage 1. The 35K stage and the 3K stage are cooled by two stages of the cryocooler. The sorption fridge is mounted on and precooled by the Stage 2. In the bottom assembly, we have used bars made of G-10 and knuckles made of aluminum glued with Stycast Epoxy to mount Stage 1 on the bottom of the cryostat, and then Stage 2 on top of Stage 1. We have also used a G-10 post to mount the chip holder on Stage 2. The heat load and the imperfect connection with the cold head of the sorption fridge allowed the sample stage to reach a base temperature of 1.7 K.

**Table 1 Heat load ($Q_c$) budget of the parts connecting stages at different temperatures.**

|  | Material | Temperatures gradient | $Q_c$ |
|---|---|---|---|
| Top Assembly |  |  |  |
| Hangers | G-10 | 300 K – 35 K | 0.71 W |
| Hangers | G-10 | 35 K – 3 K | 0.32 W |
|  |  |  |  |
| Bottom Assembly |  |  |  |
| Knuckled bars | G-10/aluminum | 300 K – 35 K | < 1.0 W |
| Knuckled bars | G-10/aluminum | 35 K – 3 K | < 88 mW |
| Chip mount holder | G-10 | 3 K – 0.8 K | 1.8 mW |

We estimated the contribution to $Q_c$ from thermal radiation [17]. We calculated that the heat radiated from the room temperature stage to Stage 1 is ~ 2.5 W. We therefore constructed a radiation shield mounted directly to Stage 1 both at the top and at the bottom of the cryostat. We cut a window in the radiation shield to allow the light from Source 2 to access the cryostat; for this window, we used a visible light filter to block parasitic radiation (i.e., not coming from our source). In addition to the radiation shield, we used a 10 layer superinsulation made of aluminum-coated Mylar between the radiation shield and the outer walls (at room temperature), between the radiation shield and Stage 2, and around the copper braid connecting the cold head of the sorption fridge to the chip holder. Each layer of Mylar is expected to reduce the amount of incident heat due to radiation up to a factor 2.

**4. Mechanical vibrations**

Another requirement for our cryostat was to reduce the mechanical vibrations between the optical system and the SNSPD below 3 μm. Mechanical vibrations with amplitude > 3 μm can reduce the average coupling efficiency of the optical set-up by more than 10%. When the Cryomech pulse-tube was operated without modification, the 3K stage vibrated with an RMS amplitude of 10 μm. As shown in Fig. 3a, the cryocooler was rigidly fixed at the top to the external chassis of the cryostat and to the optical table. Because of the large mass of the entire system (600 kg including the optical table) the vibrations were damped at the top of the cryostat. However, the cryocooler acted like a vertical cantilever, so that the two pulse-tube stages still vibrated. We further isolated the vibrations by using a combination of OFHC copper bars and braids to connect the pulse-tube stages to Stage 1 and Stage 2 at the top and bottom of the cryostat. The copper bars guaranteed high heat conduction, while the soft braids decoupled the detector from the cantilevered vibrations from the pulse-tube.

We determined that the amplitude of the vibrations at the sample stage was 498 ± 98 nm by using the time-dependence of the count rate from the detector. Fig. 4a shows the count rate measured on an 8 × 7.3 μm$^2$ area NbN SNSPD, when the beam was shifted along the longer axis of the SNSPD by 4.6 ± 0.5 μm from the center of the detector. When the center of the beam and the center of the detector were coincident, we could not see the same oscillations because of the small effect on the light coupling. On the edge of the detector, a change in the beam position produced a larger relative change in the impinging power. Fig. 4b shows the Fourier Transform of the signal on Fig. 4a. A peak was present at a frequency of 1.5 Hz, which was the frequency of the piston of the pulse-tube's engine. Fig. 4c shows the average and the standard deviation (error bar) of the count rate measured as a function of the distance between the beam's center and the SNSPD's center. The curve was fitted to extrapolate a beam waist of 6.6 ± 0.11 µm. From the beam waist and the standard deviation of the count rate, we determined a vibration amplitude of 498 ± 98 nm at 1.5 Hz. We measured that the beam power changed with a standard deviation of 5.25%, and we included it in the vibration amplitude error.

We measured the vibrations of the sample stage at higher frequencies by using the oscillations in the power of the light reflected by a highly reflective chip, shown in Fig. 4d. We focused the infrared beam on the edge of a Si chip coated with a 50 nm Au layer mounted on the sample holder. The light reflected by the chip was focused on a fast free-space photodetector (bandwidth DC – 460 kHz), which was connected to an oscilloscope. Fig. 4f shows the reflected power as a function of the beam position from the edge of the chip. By fitting the curve in Fig. 4f, we determined that the beam waist was $w = 7.93$ µm. We were not able to perform optimal focusing because we were not using an SNSPD for fine alignment. From the oscillations traces we observed vibrations at ~ 1.5 Hz and 19 Hz with amplitudes of 170 ± 50 nm and 91 ± 50 nm, respectively, which can be seen also in the FFT graph in Fig. 4e. Although this measurement shows a lower vibration amplitude compared to the measurement described in the previous paragraph, the frequency is the same that we observed in the count rate measurement. We were not able to explain the origin of the vibrations observed at 19 Hz. We did not observe oscillations at higher frequencies.

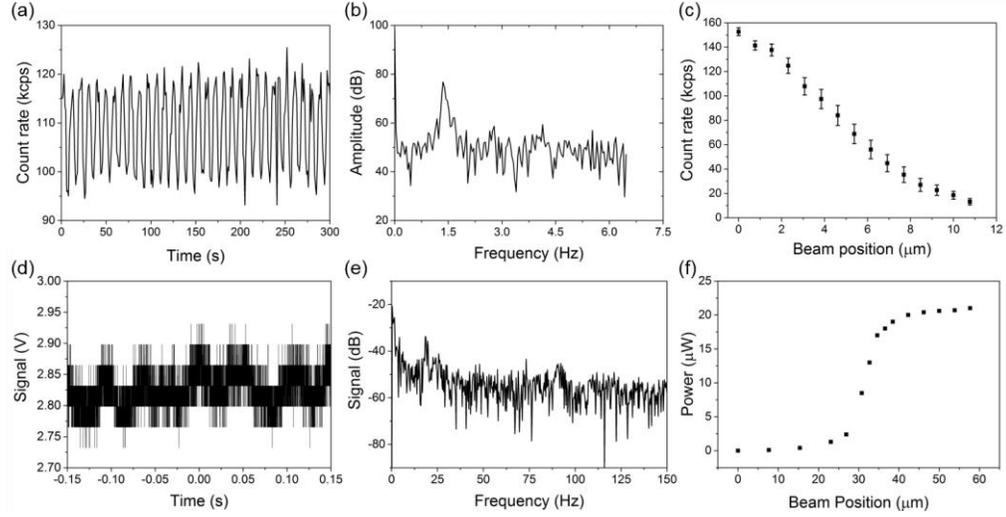

Fig. 4. (a) Count rate from an 8×7.3 μm² area NbN SNSPD as a function of time. The signal beam was positioned at 4.6 μm from the center of the detector. (b) Fast Fourier Transform of the signal shown in (a). The peak at 1.5 Hz is at the same operating frequency of the pulse-tube. (c) Count rate from the SNSPD as a function of the distance between the beam center and the SNSPD center. (d) Oscilloscope trace of the light power reflected by the edge of a gold-coated chip. (e) Fast Fourier Transform of the signal shown in d. We were able to identify the origin of the peak at 19 Hz, yet. (f) Light power reflected at the edge of a gold-coated chip as a function of the beam position.

## 5. Free-space coupling demonstration

In Section 4 we proved that we could couple near-IR light with free-space optics on an 8 × 7.3 μm² area NbN SNSPD based on 100 nm wide nanowires. We tested the *SDE* using the same detector. The *SDE* is defined as the ratio between the photon count rate (PCR) registered by the detector (excluding dark counts) and the photon flux measured in fiber at the optical source. The count rate measured when the optical source is off is the system dark count rate (*SDCR*).

In Fig. 5a, we plotted *SDCR* (blue squares) and *PCR* (red triangles) as a function of the bias current, $I_{bias}$, applied to the detector normalized by its switching current ($I_{sw}$). As we can see from the graph, *SDCR* < *PCR* for $I_{bias}$ up to 97% of $I_{sw}$ at 1.7 K. Thus, we can reliably extract *PCR* from the total counts and the *SDE*, which we plotted in Fig. 5b as a function of

$I_{bias}$ applied to the detector normalized by $I_{sw}$. The system reached a maximum *SDE* of 1.64% ± 0.13%. For this test, we used a coherent light source with $\lambda$ = 1.55 µm with a power of 710 ± 37 fW measured in fiber outside the cryostat, which corresponds to a total photon rate of 5.53 ± 0.29 Mphoton/s. From the beam waist measurement described in Section 4 ($w$ = 6.6 ± 0.11 µm) and the active area of the detector, we calculated that the coupling efficiency (*CE*) of the system was 56.5% ± 3.4%. Thus, we can calculate the maximum device detection efficiency *DDE* = *SDE/CE* = 2.9%.

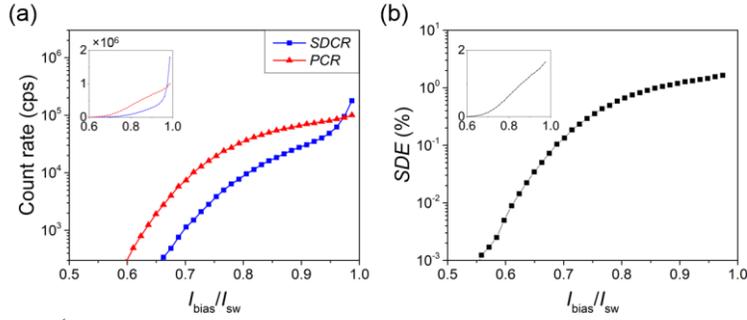

Fig. 5. (a) System dark count rate (*SDCR*, blue squares) and photon count rate (*PCR*, red triangles) as a function of the bias current normalized by the switching current of the SNSPD. The *SDCR* is determined by measuring the count rate of the detector while the source is turned off; no other filter is applied to the optical system. The *PCR* is determined by measuring the count rate when the optical source is turned on and by reducing it by the *SDCR*. (b) System detection efficiency (*SDE*) and device detection efficiency (*DDE*) as a function of the bias current normalized by the switching current of the SNSPD. The *SDE* is determined as the photon count rate divided by the photon emission rate of the source.

## 6. Discussion

The scope of this work was to create a free-space coupled cryogenic system for the use of SNSPDs in the mid-IR dispersive optics QKD communication. The use of free-space optics allows us to adapt the optical set-up for a different wavelength range by replacing the lenses. Our demonstration showed promising results of obtaining *SDE* > 50%. In particular, if we were to use a 10×10 µm² active area detector, we would achieve a *CE* = 76%, and the device detection efficiency would be the only limitation to a receiver with *SDE* > 70%. In addition, our measurements showed that we can reliably characterize the *DDE* of an SNSPD.

We could improve the performance of the cryostat if we were to use better quality copper braids and if we were to mount the chip directly on the cold head of the sorption fridge. The OFHC copper braids that we use at the moment are soldered with silver based solder to copper plates for easy mounting. If we were to e-beam weld those parts instead, we could obtain a conductivity twice as high as the present value. In addition, mounting the chip directly to the sorption fridge would reduce the temperature gradient between the two. This change requires a substantial redesign of the cryostat.

Our next step will be to repeat the demonstration with mid-IR optics at $\lambda = 10$ μm, which will require a system with a larger numerical aperture and wavelength filters. For future experiments at 10-μm-wavelength, we will replace the optical components with materials compatible with the mid-IR, such as germanium or zinc-selenide. In addition, we will replace Lens 1 with a larger focal length lens, while keeping Lens 3 at the same focal length, because we will need a stronger demagnification at mid-IR wavelength. We estimate that losses due to these components to be around 9% due to reflections. Furthermore, we will replace the CCD camera with an IR camera to image the SNSPD chip. We calculated that even if we were to use a spatial filter on the radiation shield to cut off thermal radiation at $T = 300$ K, we could still observe $> 10^6$ cps due to stray radiation. Thus, we are designing a bandpass filter system to reduce the thermal radiation bandwidth to the minimum. Once we can prove high coupling efficiency at the mid-IR wavelength, we will attempt to couple multiple sources to an SNSPD array.

Finally, we will operate SNSPDs based on WSi, instead of NbN. The reason for this consideration is that WSi is a material with a lower bandgap that NbN, so it is more sensitive to low energy photons. As the bandgap in WSi is lower than in NbN, proportionally the critical temperature of WSi is lower. A temperature $T < 2$ K is required to operate thin-film WSi SNSPDs.

**7. Conclusion**

Our system represents the first step in the realization of a mid-IR single-photon receiver for free-space optical communication. Upgrading this receiver to a multi-channel system can allow single-photon communication at 100 Mbit/s. This technology could not only allow low-power secured maritime communications, but it could also be applied to space-to-earth communication.

Outside of the secured optical communication framework, our free-space coupled system could be used for several other applications. A version of our system with visible-wavelength optics could be used to study the time-resolved emission of NV-centers, which are a key technology for quantum photonics. Mid-IR spectroscopy has become important for gas-sensing, and we could use a multi-channel version of our system to study the evolution of µs-time-scale chemical reactions.

**Acknowledgement**

The authors thank J. Daley, M. Mondol, and Dr. V. Anant for technical support and Dr. J. Mower, Dr. F. Marsili, and Dr. Y. Ivry. This work was sponsored by the Office of Naval Research ONR BAA 13-001 and by the Intelligence Advanced Research Projects Activity (IARPA) via Air Force Research Laboratory (AFRL) contract number FA8650-11-C7-105. A.E.D. and A.M.C. were supported by the iQuISE fellowship.

**References and links**